# Gaussian modulated hyperbolic tangent high pass filter for edge detection in noisy images

S.Anand, G Sangeethapriya
*Department of Electronics and Communication Engineering*
*Mepco Schlenk Engineering College, Sivakasi, India*

**Abstract**

In this paper, a non-separable (NS), robust to noise, Two Dimensional (2D) isotropic Gaussian Modulated Hyperbolic Tangent (GMHPT) High Pass (HP) filter is designed to filter the high frequency components present in a noisy image. The major drawbacks of conventional HP filters used for image analysis are noise influences (more sensitivity to noise) which lead to spurious responses and limited directional selectivity due to their separable property. The designed filter employs GMHBT frequency response to ensure improved noise performance in the least square sense. The 2D NS filter has better directional selectivity and GMHBT profile offers less noise sensitivity along with regularization by least square error design. Improvement of high pass filtering in noisy images, by means of restoring the high frequency components of the image is measured with Peak Signal to Noise Ratio (PSNR) and compared with two dimensional, non-separable isotropic Laplacian of Gaussian (LoG) filter. The designed filter gives 3dB improvement in the PSNR.

*Keywords: 2D Non Separable, High Pass filter, Least Square error, hyperbolic tangent*

## 1. Introduction

High Pass (HP) filtering is an important procedure of computer vision in which objects need to be recognized by their edges [1] and have different applications [2] [3]. In an image, the frequency content due to edges concentrated at the right end of the spectrum. HP filtering is the common approach used to determine the amount of gray level change present in an image at every pixel. The gray level change detectors generally based on the image gradient and the most familiar Sobel edge filter is the mean weighted HP filter, which captures a weighted difference between the pixels on either side and performs better than the other edge operator like Roberts, and Prewitt. However the Sobel filter is separable, anisotropic, and normally suffers from double edges. Due to this separable property, they have poor directional selectivity. The detection of edge pixels is based on differential components and hence more assert to noise. Therefore to improve the directional selectivity and the noise performance a Two Dimensional (2D) non separable (NS) filter is designed based on the constrained least square error sense with GMHBT circular frequency samples.

2D NS filters are used to improve more directional features [4, 5]. The potential application of hyperbolic tangent filters is well understood in [6]. The properties of these filters are discussed and its improved noise performance in edge detection also given in [7]. The least square filter designs for better regularization are designed in [8] and an edge extraction filter also found [6] in the literature. Various approaches to designing NS filters are proposed in [10, 11]. The common gradient based operators like Sobel, Roberts, Prewitt's, etc. have a separable implementation, whereas Canny uses the differences of Gaussian for detecting edges. Laplacian of Gaussian (LoG) filter operator has two dimensional NS filter masks and its noise influence is removed by fine-tuning the smoothing parameter [12]. But this leveling process fades the genuine gradient and this is a concession between the smoothing parameter and the number of pixels (edge) spotted is required [13].

## 2. Least Square Two Dimensional Filter Design

2D Finite impulse response (FIR) filters have many preferable features that mark them perfect for computer vision [14]. Some of them are: 1). FIR filters are easy to represent as matrices of coefficients. 2). several reliable methods are available for 2D FIR filter design. 3). FIR filter coefficients are obtained to have zero or linear phase response, which is an important property in many imaging applications to stop alteration [15]. The basic prerequisite enforced in this filter is the sum of all the impulse-response values should be equivalent to unity to avoid bias in the light intensity through the handling. This requirement ensures that the image background will not be affected by the noise removal process.

A 2D non recursive filter can be represented by

$$y(x_1, x_2) = \sum_{k_1=-C_1}^{C_1} \sum_{k_2=-C_2}^{C_2} h(x_1 - k_1, x_2 - k_2) I(k_1, k_2) \tag{1}$$

where $C_1 = \frac{x_1-1}{2}, C_2 = \frac{x_2-1}{2}$, $h(x_1, x_2)$ is the impulse response and $I(k_1, k_2)$ is the image to be processed, of order $N_1 = N_2 = N$ where $N_1$ and $N_2$ are odd. The filter satisfies the condition $\sum_{x_1=-C_1}^{C_1} \sum_{x_2=-C_2}^{C_2} h(x_1, x_2) = 1$ to have zero phases. An HP zero phase circularly symmetric frequency response is expressed as

$$H(\omega_1, \omega_2) = \frac{(\omega_1^2 + \omega_2^2)^N}{0.4142\omega_c^{2N} + (\omega_1^2 + \omega_2^2)} \tag{2}$$

where $\omega_c$ is the cutoff frequency. Let the filtered image $Y(\omega_1, \omega_2)$ can be obtained by

$$Y(\omega_1, \omega_2) = H'(\omega_1, \omega_2)[G(\omega_1, \omega_2) X(\omega_1, \omega_2)\},$$

where $H'(\omega_1, \omega_2) = H(\omega_1, \omega_2) G(\omega_1, \omega_2)$ and $G(\omega_1, \omega_2)$ is defined as the frequency response of Gaussian modulated hyperbolic tangent function and its spatial domain 2D representation becomes

$$G_w(x_1, x_2) = G(x_1, x_2) \cdot \begin{cases} \frac{1-e^{\sigma w(x_1+x_2)}}{1+e^{\sigma w(x_1+x_2)}} & \text{for } |x|, |y| \leq w \\ 0 & \text{otherwise} \end{cases} \quad (3)$$

The region of support is limited by '$w$' to ensure frequency localization. The filter $G_w(x_1, x_2)$ is odd symmetric with single zero crossings at the origin with the slope $\sigma_w/2$ at zero crossing point [7]. The slope parameter controls the smoothness of the transfer function around the cutoff frequency. Fig.1. compares the smoothness of the GMHBT function with the Gaussian function.

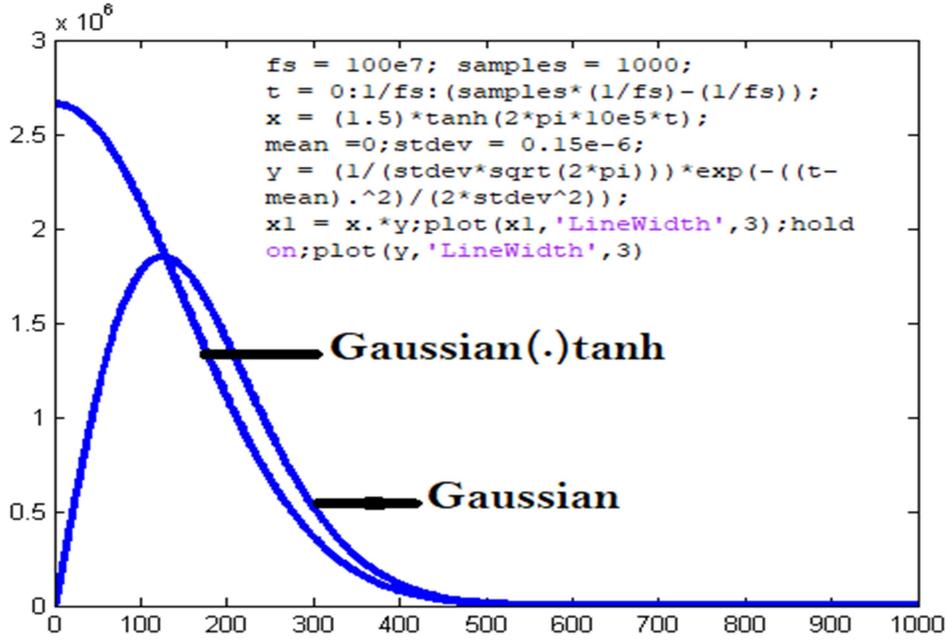

Fig.1. Smoothness parameter of GMHBT function with Gaussian function

So, special care is needed in the selection of the smoothness parameter. If this parameter takes very small values, then $G(\omega)$ becomes sn ideal filter. When the smoothness parameter takes relatively high values, the slope of filter function in the frequency domain decreases and significantly reduces the oscillations of the filter response function in the time domain. This permits the construction of a relatively short length filter. In the optimal design of $G(\omega)$, preferred frequency response is carefully approached by optimally diminishing the error. In other words, it is the problem of minimization of the $L_2$ frequency domain error in the pass band region and the stop band region while keeping an anticipated frequency response. Based on the development and simplification of [5], the optimal 2D FIR filter is considered.

Let a favorite function denoted as $D(\omega_1, \omega_2)$ and an error function be described as $E'(\omega_1, \omega_2)$

$$E'(\omega_1, \omega_2) = G'(\omega_1, \omega_2) - D'(\omega_1, \omega_2) \quad (4)$$

In its place of minimizing the weighted integral error after the above equation, $E^2(\omega_1, \omega_2)$ is diminished on the lattice $(\omega_m, \omega_n), m = 1,2 \ldots M, n = 1,2 \ldots K$ where M and K are positioned in the frequency lattice. The error function (4) can be rewritten in vector method as

$$\mathbf{E'} = \mathbf{G'} - \mathbf{D'} = \mathbf{Ag} - \mathbf{D'} \qquad (5)$$

where **E'** is the error vector, **G'** is the actual frequency response, and **D'** is the desired response and the responses are chosen in frequency grids.

The well-known solution to minimize the least square error of (5)

$$\varepsilon(g) = \mathbf{E'^T W E} \text{ is of the form}$$

$$g = \mathbf{R^{-1} d}$$

where

$$\mathbf{R} = \mathbf{A^T W A} \text{ and}$$

$$\mathbf{d} = \mathbf{A^T W D'}$$

**W** denotes the weighting matrix and matrix **A** contains real terms [8]. In the above least square inference, the constraints upper and lower bound tolerance **B** denoting the approximation frequency range are imposed and the design problem can be formulated as

$$\text{Subjected to } \mathbf{B}(g) = 0; \text{ to minimize } \varepsilon(g) \qquad (6)$$

and the Least Square solution is computed. The main characteristics of the filter can be summarized as: 1). Different smoothness parameters can be chosen for cut off frequencies, 2). Time domain filter function can be derived analytically from frequency domain and 3). Permit relatively short filter.

## 3. Experiment Results and discussion

The noise robustness of the proposed method compared to the NS LoG filter. Being the second derivative, the influence of noise is considerable in LoG and it generates the closed contour. The noise effects minimized by smoothing with a Gaussian shaped kernel before applying the Laplacian operator. Laplacian operator subtracts the brightness values of each of the neighboring pixels from the central pixel. When the high frequency component in the form of discontinuity is present within the neighborhood, the result of the Laplacian is nonzero value. Laplacian operator is rotationally invariant that is, it does not depend on directions as long as they are orthogonal, and being an approximation to the second derivative, it enhances noises.

An HP filter is designed as in section 2, and the obtained results are shown in Fig.2.-Fig.4. The filter is designed for the length N = 5, with $\sigma_w = 0.7$, and the obtained filter coefficients are given in Table I. From the HF components, the edge pixels are easily recognized by applying threshold technique. Fig.2. (a), (b), and (c) shows the original *Elaine* image from USC-SIPI image database with a size of 512 x 512, applied with the designed filter, and LoG filter respectively. Both filter Fig.2. (b) and, (c) yields nearly the same result.

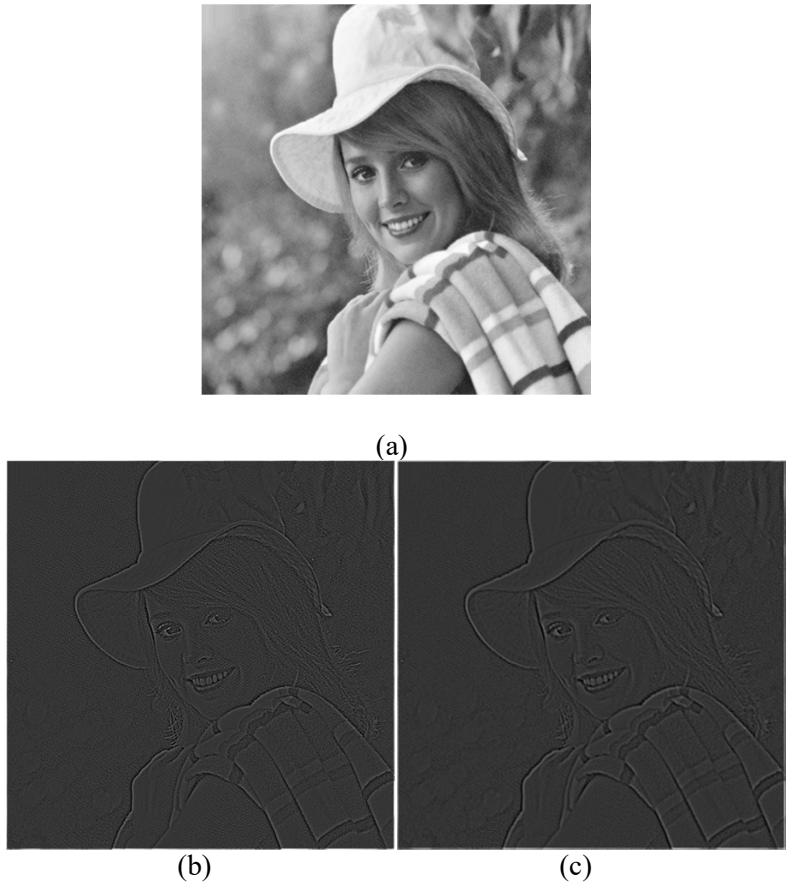

Fig.2.High frequency components of Elaine Image (a) Original (b) HP filtered by LoG filter

(c) HP filtered by Designed filter

### 3.1. Noise Environment:

Since noise is the high-frequency content, the operators used on noisy images are typically larger in scope, so they can average enough data to discount localized noisy pixels. The filter averages a specified number of pixels on either side, known to be the length or size of the filter. Longer filters average more points and are less susceptible to identifying noise and hence less accurate in localization. The actual location is a matter of interpretation and is somewhat application-dependent. Our algorithm returns the high frequency components, halfway between the two crossings of the contrast threshold. For the experimental purpose, the length of the filter is taken as 3 and 5.

It is also observed that this filtering scheme effectively yields HF components in the noisy environment from Fig. 3(b). On the contrary, the sensitivity to noise of the LoG technique is more than the designed method as in Fig. 3(c). In order to obtain a quantitative evaluation of the noise performance, we considered a set of 6 pictures taken from USC-SIPI image database and corrupted by Gaussian noise with a noise variance of $\sigma^2 = 25$ and $\sigma^2 = 15$. The noise performances are measured by adopting the approach proposed in [12], where the HF maps of the images corrupted by noise are compared with the HF maps of the original image for each method. The corresponding noise performance is measured by means of the peak signal-to-noise ratio (PSNR), which is defined as

$$\text{PSNR} = 10 \log_{10}\left(\frac{\sum_i \sum_j (L-1)^2}{\sum_i \sum_j (y(i,j) - z(i,j))^2}\right)$$

where L is the number of luminance levels in an image (i.e.) taken as 256 in our experiment, $y(i,j)$ is the pixel luminance at location $(i,j)$ in the high frequency map of the noisy image and $z(i,j)$ denotes the corresponding pixel luminance in the high frequency map of the original picture.

The PSNR values of the restoration process of LoG and proposed methods for various images are reported in Table II. The performance improvement given by the proposed method in terms of restoring HF maps from noisy conditions is very relevant (at least 4 dB with respect to LoG technique) and satisfactory compromise in noise sensitivity is achieved. From the Table II, it is also studied that, when the noise level increase, this filter is superior to the LoG filter, (i.e.) when the noise variance varies from 15 to 25.

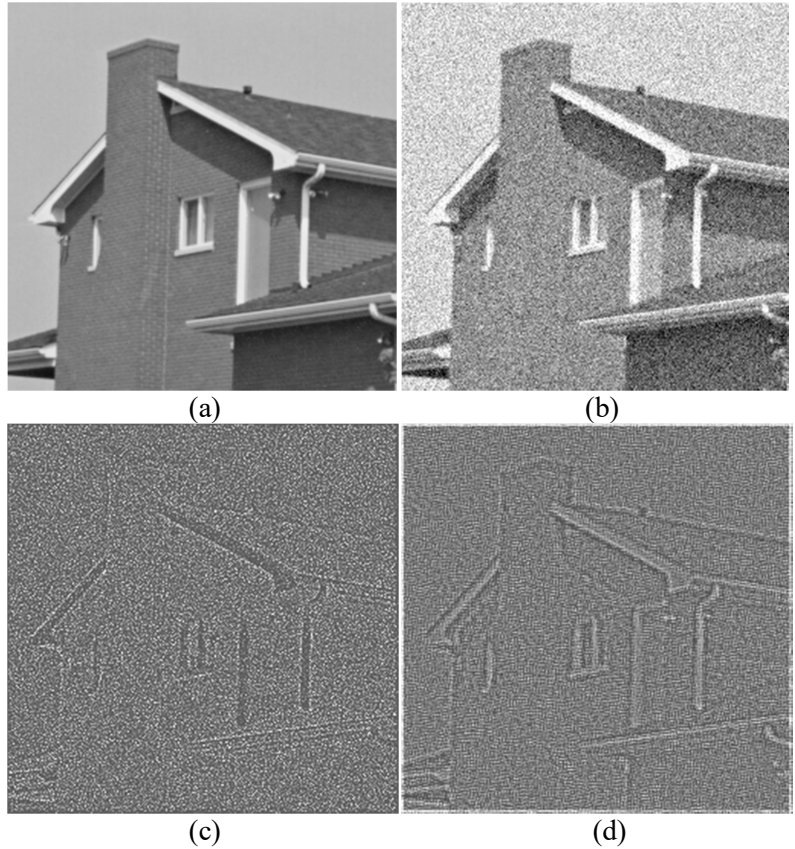

Fig.3.Restoration of High frequency components from noisy House Image (a) Original (b) Corrupted by Gaussian noise ($\sigma^2 = 25$) (c) LoG filter (d) Designed filter

TABLE I

Filter Coefficients of 2D NS Filter with N=5

| $h'(n_1,n_2)$ | $n_1$ | $n_2$ | $n_3$ | $n_4$ | $n_5$ |
|---|---|---|---|---|---|
| $n_1$ | 0.8854 | 0.7818 | 0.6044 | 0.7818 | 0.8854 |
| $n_2$ | 0.7818 | 0.6044 | 0.3364 | 0.6044 | 0.7818 |
| $n_3$ | 0.6044 | 0.3364 | 0 | 0.3364 | 0.6044 |
| $n_4$ | 0.7818 | 0.6044 | 0.3364 | 0.6044 | 0.7818 |
| $n_5$ | 0.8854 | 0.7818 | 0.6044 | 0.7818 | 0.8854 |

TABLE II

Performance Comparison of Proposed Filter: Restoration process of High frequency components from the noise environment

| Image | Filter | PSNR in dB | | | |
|---|---|---|---|---|---|
| | | Filter Size = 5 x 5 | | Filter Size = 7 x 7 | |
| | | $\sigma = 10$ | $\sigma = 20$ | $\sigma = 10$ | $\sigma = 20$ |
| House | LoG | 9.28 | 4.88 | 9..23 | 4.82 |
| | Proposed | 14.07 | 12.35 | 16.27 | 15.77 |
| Lena | LoG | 11.26 | 4.83 | 11.24 | 4.83 |
| | Proposed | 16.30 | 13.70 | 20.15 | 20.08 |
| Boat | LoG | 10.39 | 5.95 | 10.32 | 4.80 |
| | Proposed | 14.36 | 12.50 | 18.30 | 17.56 |
| Bridge | LoG | 10.38 | 5.95 | 10.54 | 4.77 |
| | Proposed | 12.52 | 11.33 | 17.57 | 16.76 |
| Gray21 | LoG | 10.28 | 4.82 | 09.44 | 4.85 |
| | Proposed | 16.59 | 14.04 | 19.75 | 18.76 |
| Elaine | LoG | 9.28 | 47 | 10.35 | 7.73 |
| | Proposed | 16.35 | 13.90 | 19.97 | 18.91 |

**3.2. Image Sharpening Application**

Image sharpening is one of the image enhancement methods, which improve the contrast around the edges of objects to increase the image's definition [16] and can apply to an image to bring out image detail that was not there before. However, it is to emphasize edges and make the image appear sharper. While sharpening an image, two effects occur [17]; 1). Edges become unnaturally pronounced (i.e.) dark objects have light radiance outline and light objects have dark radiance outlines. 2). Invisible noise in the image is amplified and texture in areas that looked smooth in the original images. The above difficulties are efficiently avoided in the image sharpening method, which uses the proposed filter result. The HP result of our filter is superimposed with the original to get the enhanced image. It is well evidenced in Fig. 4(c) texture region of *Lena* image cap.

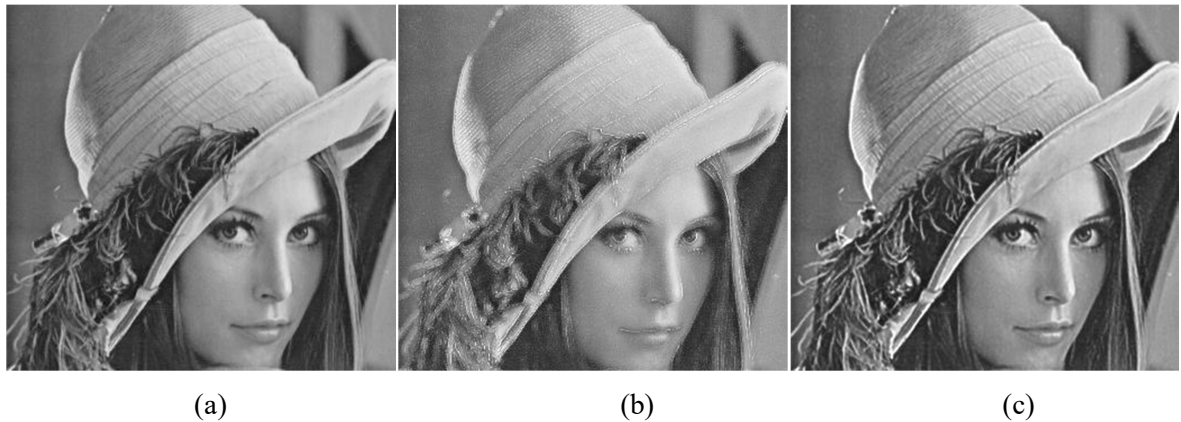

(a)            (b)            (c)

Fig.3. Image Sharpening of Lena Image (a) Original (b) Image Sharpening using LoG
(c) Image Sharpening using the proposed filter

### 4. Conclusions

In this paper, GMHBT based HP filter is designed to filter the high frequency for noisy images that have been presented. The proposed least square approach makes use of noise performance GMHBT profile in the 2D NS filter design and yields HF components better than LoG. This 2D NS filter improves directional selectivity, while the GMHBT provides less noise sensitivity and regularity by least square design. PSNR is used to measure the effective restoration of high frequency components in noisy images. The performance of the designed filter is compared with that of LoG for natural and synthetic images and gives robustness in extracting HF features.

**References**

[1]. Anand, S., and R. Shantha Selva Kumari. "Sharpening enhancement of computed tomography (CT) images using hyperbolic secant square filter." Optik 124.15 (2013): 2121-2124.

[2]. Anand, S., and NM Mary Sindhuja. "Spot edge detection in microarray images using balanced GHM multiwavelet." 2009 International Conference on Control, Automation, Communication and Energy Conservation. IEEE, 2009.

[3]. Muthukarthigadevi, R., and S. Anand. "Detection of architectural distortion in mammogram image using wavelet transform." 2013 International Conference on Information Communication and Embedded Systems (ICICES). IEEE, 2013.

[4]. Minh N. Do and Martin Vetterli, The Contourlet Transform: An Efficient Directional Multiresolution Image Representation, IEEE Transactions on Image Processing, 14 (12) (2005) 2091–2106.

[5]. Yuichi Tanaka, Masaaki Ikehara, and Truong Q. Nguyen, Multiresolution Image Representation Using Combined 2-D and 1-D Directional Filter Banks, IEEE Transactions on Image Processing, 18 (2) (2009) 269-280.

[6]. Ahmet T. Basokur, Digital filter design using the hyperbolic tangent functions, Journal of the Balkan Geophysical Society, 1 (1) (1998) 14-18.

[7]. Saravana Kumar, Sim Heng Ong, Surendra Ranganath, and Fook Tim Chew, A Luminance- and Contrast-Invariant Edge-Similarity Measure, IEEE Transactions on Pattern Analysis and Machine Intelligence, 28 (12) (2006) 2042-2048.

[8]. Markus Lang, Ivan W. Selesnick, and C. Sidney Bums, Constrained Least Squares Design of 2-D FIR Filters, IEEE Transactions on Signal Processing, 44 (5) (1996) 1234-1241.

[9]. Sanjit K.Mitra and Giovanni L.Sicuranza, Nonlinear Image Processing, Academic Press, (2001).

[10]. Anamitra Makur, V.L. Narayana Murthy, Design of two-dimensional directional filters, Signal Processing 60 (1997) 315-324.

[11]. Zuo-feng Zhou, Yuan-yuan Cheng, and Peng-lang Shui, Construction of 2-D directional filter bank by cascading checkerboard-shaped filter pair and CMFB, Signal Processing 88 (2008) 2500–2510.

[12]. Fabrizio Russo, and Annarita Lazzari, Color Edge Detection in Presence of Gaussian Noise Using Nonlinear Pre filtering IEEE Transactions on Instrumentation and Measurement, 54 (1) (2005) 352-358.

[13]. Rishi R. Rakesh, Probal Chaudhuri, and C. A. Murthy, Thresholding in Edge Detection: A Statistical Approach, IEEE Transactions on Image Processing, 13 (7) (2004) 927-936.

[14]. Wu-Sheng Lu, Andreas Antoniou, Two-Dimensional Digital Filters, Marcel Dekker, Inc. New York**,** 1992.

[15]. Lim, Jae S., Two-Dimensional Signal and Image Processing, Englewood Cliffs, NJ, Prentice Hall, 1990.

[16]. Konstantinos Konstantinides,Vasudev Bhaskaran, and Giordano Beretta, Image Sharpening in the JPEG Domain, IEEE Transactions on Image Processing, 8 (6) (1999).

[17]. Jonathan Sachs, Sharpening Images, http://www.dl-c.com/ (18.09.2010)